\documentclass[vecphys]{svmult}

\begin{document}

\title*{QCD corrections to Higgs physics at the LHC}
\author{Giuseppe Bozzi}
\institute{Laboratoire de Physique Subatomique et de Cosmologie,\\
Universit\'e Joseph Fourier/CNRS-IN2P3,\\ 
53 Avenue des Martyrs, F-38026 Grenoble, France\\
\texttt{bozzi@lpsc.in2p3.fr}}
\maketitle

We summarize the status of QCD corrections for SM Higgs boson production at hadron colliders and briefly sketch the main search strategies at the LHC. 

\section{Introduction}
\label{sec:1}
The Higgs boson\cite{Gunion:1989we}, which is responsible for the electroweak symmetry breaking in the Standard Model, is yet to be discovered. The experimental searches performed at LEP allowed to put a lower bound $m_{H} > 114$ GeV  \cite{Barate:2003sz} on the mass $m_H$ of the SM Higgs boson, whereas an upper limit  $m_{H} < 219$~GeV at $95\%$ CL has been obtained through global SM fits to electroweak precision measurements \cite{Group:2005di}. 

The search for the Higgs boson will be one of the most important tasks for the CERN LHC \cite{atlascms,lhcupdate}. A considerable effort has been devoted in recent years to improve the theoretical predictions: the most relevant production mechanisms and decay modes of the SM Higgs boson, together with the main background processes, are now known to high precision.

This brief overview is mainly intended to summarize the status of the QCD corrections to SM Higgs boson production at the LHC, focusing on the gluon fusion and weak boson fusion processes. For a comprehensive review, we refer the reader to \cite{Djouadi:2005gi}.

\section{SM Higgs production at the LHC}
\label{sec:2}
\textbf{Gluon fusion.} At the LHC, the Higgs boson will be predominantly produced through the gluon fusion process $gg \rightarrow H$, involving a heavy-quark (mainly top-quark) loop. The total cross section at leading-order (LO) in QCD perturbation theory (${\cal O}(\alpha_{s}^{2})$) has been known for a long time \cite{Georgi:1977gs}. The next-to-leading order (NLO) corrections \cite{Dawson:1991zj, Djouadi:1991tk, Spira:1995rr}, are very large (LO is increased by up to 100\%). The complexity involved with the heavy-quark loop makes the computation of higher-order corrections extremely difficult. In the large-$m_{t}$ limit ($m_{t} \gg m_{H}$) it is possible to introduce an effective lagrangian \cite{Ellis:1975ap} that directly couples the Higgs to gluons:
\begin{equation}
\label{eff}
{\cal L}_{\rm eff} =  -\frac{1}{4} [1-\frac{\alpha_{s}}{3\pi}\frac{H}{v}(1+{\rm \Delta}) ] \ {\rm Tr} \ {\cal G}_{\mu\nu} {\cal G}^{\mu\nu}
\end{equation}
where the coefficient $\rm \Delta$ is known up to ${\cal O}(\alpha_{s}^{3})$ \cite{Chetyrkin:1997iv}. It was shown~\cite{Kramer:1996iq} that NLO calculations based on the effective lagrangian approximate the full NLO result within 10\% up to $m_{H}$=1 TeV. The high accuracy of this approximation is due to the fact that the Higgs boson is predominantly produced in association with partons of relatively low transverse-momenta, which are unable to resolve the heavy-quark loop \cite{Catani:2001ic}.
The next-to-next-to-leading order (NNLO) corrections have been evaluated in the large-$m_{t}$ limit \cite{Catani:2001ic,Harlander:2000mg,Harlander:2001is,NNLOtotal}. In the case of a light Higgs boson ($m_{H}\sim$ 100-200 GeV), the total cross section increases by about 10-25\% at LHC and shows a reduced scale dependence (10-15\%), thus exhibiting the features of a well-behaved perturbative series. In addition to the fixed-order results, the resummation of logarithmically-enhanced terms due to multiple soft-gluon emission has been carried out at the full next-to-next-to-logarithmic (NNLL) level \cite{Catani:2003zt} and at the ${\rm N^{3}LL}$ level \cite{Moch:2005ky}, providing a further 7-8\% increase w.r.t. NNLO and reducing scale dependence to less than 4\%. Also the NLO EW contributions have been recently computed \cite{Aglietti:2004nj}, showing a 5-8\% effect below the $m_{H} = 2 m_{W}$ threshold. 

Reliable theoretical predictions for the Higgs differential ($q_{T}$ and $y$) distributions are necessary to better understand the kinematics of the final states and compare with realistic experimental acceptances. The most advanced predictions available at present are the NNLO rapidity distribution \cite{Anastasiou:2005qj}, and the NLO (${\cal O}(\alpha_{s}^{4})$) transverse-momentum distribution \cite{deFlorian:1999zd}. These predictions in fixed-order perturbation theory have been supplemented with the resummation of higher-order corrections due to multiple soft-gluon emission. In the region of small transverse-momentum, $q_{T}$-resummation has been performed at the NNLL level \cite{Bozzi:2003jy}, while joint (threshold and $q_{T}$) resummation has been performed at the full NLL level \cite{Kulesza:2003wn}.\\
\\
\textbf {Weak boson fusion.} This production mechanism occurs as the scattering between two (anti)quarks with weak boson ($W$ or $Z$) exchange in the $t$-channel and with the Higgs boson radiated off the weak-boson propagator. Since the parton distribution functions of the incoming valence quarks peak at values of the momentum fractions $x\sim$ 0.1 to 0.2, this process tends to produce two highly-energetic outgoing quarks. In addition the large weak boson mass provides a natural cutoff on its propagator: as a consequence, the jets are produced with a transverse energy of the order of a fraction of the weak boson mass and thus with a large rapidity interval between them (typically one at forward and the other at backward rapidity). Moreover, since the exchanged weak boson is colourless, no further hadronic production occurs in the rapidity interval between the quark jets (except for the Higgs decay products). All these phenomenological features make Higgs production via weak boson fusion (WBF) a very promising tool for precision measurements at LHC.

The LO partonic cross section can be found in \cite{Cahn:1983ip}. As for QCD corrections, gluon radiation occurs to ${\cal O} (\alpha_{s})$ only as bremsstrahlung off the quark legs. NLO corrections to Higgs production via WBF have been computed for the total cross section~\cite{Han:1991ia} and for Higgs production in association with two jets~\cite{Figy:2003nv}. They have been found to be typically modest (5-10\%).

\section{SM Higgs searches at the LHC}
\label{sec:3}
\textbf{$m_{H} <$ 140 GeV.} In the lower mass range the Higgs particle dominantly decays into $b \bar b$ pairs. Because of the overwhelming QCD background the signal will be very difficult to extract, thus this decay mode is almost discarded. The same considerations apply to the leptonic decay $H\to\tau^{+}\tau^{-}$, hidden by the huge Drell-Yan lepton pair production background. Thus, the most promising channel at LHC in the case of a light Higgs is $gg \rightarrow H \rightarrow \gamma\gamma$. The decay $H \rightarrow \gamma\gamma$ is known up to 3-loop QCD \cite{Steinhauser:1996wy}, while the irreducible $pp \rightarrow \gamma\gamma$ background is available at NLO in the program DIPHOX \cite{Binoth:1999qq}, which also includes all relevant photon fragmentation effects. The loop-mediated process $gg \rightarrow \gamma\gamma$ contributes about 50\% to the background and has been calculated at ${\cal O}(\alpha_{s}^{2})$ \cite{Bern:2002jx}.\\
\\
\textbf{140 GeV $< m_{H} <$ 180 GeV.} For intermediate Higgs masses the relevant processes are $qq \rightarrow qqV^{*}V^{*} \rightarrow qqH, (H \rightarrow \gamma\gamma,\tau^{+}\tau^{-},V^{*}V^{*})$ and $gg \rightarrow H \rightarrow W^{+}W^{-} \rightarrow l^{+}l^{-}\nu \bar \nu$. In the case of the first (WBF) channel, the decays $H \rightarrow \tau^{+}\tau^{-},V^{*}V^{*}$ are known at 2- \cite{Djouadi:1994gf} and 3-loop \cite{Kniehl:1995br} QCD respectively. The dominant background is Higgs production through gluon fusion in association with two jets. The full top mass dependence of the cross section at LO (${\cal O}(\alpha_{s}^{4})$) has been evaluated in \cite{DelDuca:2001fn}. Other important backgrounds to WBF are $Vjj$ and $VVjj$ production, for which NLO corrections are available \cite{Oleari:2003tc}. The second (gluon-initiated) channel is the most challenging one because backgrounds are of the order of the signal rate or larger. The dominant processes are $pp \rightarrow W^{+}W^{-}$, known at NLO \cite{Baur:1997kz} and recently supplemented with soft-gluon effects \cite{Grazzini:2005vw}, and off-shell $t\bar t$ production, known only at LO \cite{Kauer:2001sp}.\\
\\
\textbf{$m_{H} >$ 180 GeV}. Above the ZZ threshold  the $H \rightarrow ZZ \rightarrow l^{+}l^{-}l^{+}l^{-}$ process provides a very clean signature with small background \cite{Gunion:1987ke}.\\
\\
The great theoretical effort made in the last years provided us with a very detailed knowledge about Higgs hadroproduction and decay. This knowledge will be essential to improve search strategies, especially in the delicate low- and intermediate-mass regions. 

If the Higgs boson exists, there is no escape route for it at the LHC.

\end{document}